\documentclass[fleqn,usenatbib]{mnras}

\usepackage{newtxtext,newtxmath}
\usepackage[section]{placeins}
\usepackage[T1]{fontenc}
\DeclareRobustCommand{\VAN}[3]{#2}
\let\VANthebibliography\thebibliography
\def\thebibliography{\DeclareRobustCommand{\VAN}[3]{##3}\VANthebibliography}
\usepackage{mathtools}
\usepackage{subcaption,graphicx}
\usepackage[normalem]{ulem}
\graphicspath{{images}}
\usepackage{amsfonts}
\usepackage{comment}

\author[Uzeirbegovic et al.]{E.
  Uzeirbegovic$^{1,2}$\thanks{e.uzeirbegovic@herts.ac.uk},  G. Martin$^{3}$, S. Kaviraj$^{1,2}$,
  R. A. Jackson$^{1,4,5}$, K. Kraljic$^{6}$, Y. Dubois$^{7}$, \newauthor
  C. Pichon$^{7,8,9}$, J. Devriendt$^{10}$, S. Peirani$^{11,12}$,
  J. Silk$^{7,10,13}$ and S. K. Yi$^{5}$\\
$^{1}$Centre for Astrophysics Research, University of Hertfordshire, Hatfield, AL10 9AB, UK\\
$^{2}$Centre of Data Innovation Research, University of Hertfordshire, Hatfield, AL10 9AB, UK\\
$^{3}$School of Physics \& Astronomy, University of Nottingham, University Park, Nottingham NG7 2RD, UK\\
$^{4}$Department of Physics and Astronomy, University of Victoria, BC V8P 5C2, Canada\\
$^{5}$Department of Astronomy and Yonsei University Observatory, Yonsei University, Seoul 03722, Republic of Korea\\
$^{6}$Aix Marseille Univ, CNRS, CNES, LAM, Marseille, France\\
$^{7}$Institut d'Astrophysique de Paris, UMR 7095, 98 bis Boulevard Arago, F-75014 Paris, France\\
$^{8}$IPhT, DRF-INP, UMR 3680, CEA, L’orme des Merisiers, Bât 774, 91191 Gif-sur-Yvette, France\\
$^{9}$Korea Institute for Advanced Study, 85 Hoegi-ro, Dongdaemun-gu, Seoul 02455, Republic of Korea\\
$^{10}$Sub-department of Astrophysics, University of Oxford, Keble Road, Oxford OX1 3RH, UK\\
$^{11}$ILANCE, CNRS – University of Tokyo International Research Laboratory, Kashiwa, Chiba 277-8582, Japan\\
$^{12}$Kavli IPMU (WPI), UTIAS, The University of Tokyo, Kashiwa, Chiba 277-8583, Japan\\
$^{13}$Department of Physics and Astronomy, The Johns Hopkins University, Baltimore, MD, USA\\
}

\title[MW satellite planes are consistent with $\Lambda$CDM]{New tools for studying planarity in galaxy satellite systems: Milky Way satellite planes are consistent with $\Lambda$CDM
}

\begin{document}

\maketitle

\begin{abstract}
We introduce a new concept -- termed `planarity' -- which aims to quantify planar structure in galaxy satellite systems without recourse to the number or thickness of planes. We use positions and velocities from the Gaia EDR3 to measure planarity in Milky Way (MW) satellites and the extent to which planes within the MW system are kinematically supported. We show that the position vectors of the MW satellites exhibit strong planarity but the velocity vectors do not, and that kinematic coherence cannot, therefore, be confirmed from current observational data. We then apply our methodology to NewHorizon, a high-resolution cosmological simulation, to compare satellite planarity in MW-like galaxies in a $\rm{\Lambda CDM}$-based model to that in the MW satellite data. We demonstrate that kinematically supported planes are common in the simulation and that the observed planarity of MW satellites is not in tension with the standard $\rm{\Lambda CDM}$ paradigm. 

%We study galaxy satellite planes by combining the \textsc{NewHorizon} cosmological simulation with positions and velocities of Milky Way (MW) satellites from the Gaia EDR3, with novel tools for describing `planarity' -- a new concept aimed at quantifying planar structure without recourse to the number or thickness of planes. We measure planarity in MW satellites and the extent to which planes are kinematically supported, and explore the consequences of satellite planes for the fidelity of the $\rm{\Lambda CDM}$ paradigm. We first explain the benefits of our tools and then employ them to analyse planar structures in 3D space to describe planarity in the MW. We show that the position vectors of the MW exhibit strong planarity but the velocity vectors do not, and that kinematic coherence cannot, therefore, be confirmed from current observational data. We apply the same tools to \textsc{NewHorizon}, in order to compare satellite planarity around mock galaxies in a $\rm{\Lambda CDM}$-based model to those calculated from MW satellite data. We show that kinematically supported planes are common in the simulation and that the planarity of MW satellites is not in tension with the standard $\rm{\Lambda CDM}$ paradigm.
\end{abstract}


\section{Introduction}
\label{intro}

Systems of satellite galaxies around massive hosts have often been studied in the context of a variety of structural properties, including (1) the existence of planes \citep[]{kroupa2005,ibata2014,cauton2017,samuel2021}, (2) alignment with a principal axis \citep[e.g.][]{west2000,plionis2003}, (3) `lopsidedness' \citep[e.g.][]{libeskind2016, pawlowski2017}, and (4) rotational support \citep[e.g.][]{metz2008, ibata2013}. These studies are often used to explore the so-called `small-scale' challenges to the standard $\rm{\Lambda CDM}$ cosmology \citep[see][for an overview]{Bullock2017} where observations at spatial scales less than $\sim1~{\rm Mpc}$ have been thought to be in tension with the predictions of the $\rm{\Lambda CDM}$ model. The usual course taken is to compare observations from the Milky Way (MW) and relatively nearby galaxies (e.g. M31) -- where accurate 3D positions and some velocity information is available -- to outputs from flexible modelling techniques such as simulations \citep[e.g.][]{2023sawala, bahl2014, samuel2021} or semi-analytical \citep[e.g.][]{cauton2017, jiang2021} models. 

The comparison is often framed as a hypothesis test of the difference between real and simulated observations, via a metric strongly tied to a measure of a primary plane. For example, given some satellite system centred on a host galaxy, \cite{samuel2021} characterise planes by calculating the root mean square height (perpendicular distance of galaxy to the plane), the minor-to-major axis ratio of the inertial tensor and orbital coherence (the deviation from the mean angular momentum direction). They then determine how uncommon the metrics calculated from physical observations are, by comparing them to the empirical distributions of the measures obtained from the forward-modelling of simulations. Similar in spirit, \cite{cauton2017} use `velocity anisotropy' (the ratio of tangential and radial velocity) as their primary metric, and compare physical observations to a semi-analytical model. They also draw conclusions about differences by comparing metrics derived from physical observations to the empirical distribution of these metrics derived from simulated observations.

Many analyses of satellite galaxy systems which touch on $\rm{\Lambda CDM}$ challenges have been contested on methodological grounds. For example, \citet{kroupa2005} were amongst the first to identify Milky Way (MW) planarity as a potential challenge to $\rm{\Lambda CDM}$, on the grounds that it was inconsistent with being drawn from an isotropic distribution\footnote{One way to create an isotropic distribution (used herein) is via rejection sampling. That is, sample points from a unit cube (uniform sampling across 3 dimensions) and only retain those points falling into a unit sphere.}. However, others \citep[e.g.][]{2004aubert, libeskind2005, zentner2005} have pointed out that $\rm{\Lambda CDM}$ simulations do not yield isotropically distributed galaxy satellites in the first place. \cite{ibata2014} consider SDSS host galaxies at $z<0.5$ with diametrically opposite satellite galaxies and ask whether their line of sight velocities are typically in opposite directions, as might be expected given a rotationally supported disc. They report that this is the case for 20 out of 22 such pairs within an opening angle of $\le 8$ degrees. \cite{phillips2015} replicate this analysis confirming the main result but then go on to show that simply widening the opening angle leads to statistical insignificance and that the results achieved at $\le 10$ degrees is consistent with the under-sampling of an underlying isotropic velocity distribution. 

In a similar vein, \cite{pawlowski2014} argue that satellite planes only appear common in $\rm{\Lambda CDM}$ simulations because they fail to take into account rotational support and that, once they do, the incidence of planes is much lower. However, \cite{cautun2015} note that \cite{pawlowski2014} do not themselves take into account the fact that they are multiple hypothesis testing, and that resultantly they overestimate the significance of MW planes by around a factor of 30. More recently, \cite{hammer2021} -- utilising proper motion data of more than 40 MW dwarf satellites from the Gaia EDR3 -- show that the MW dwarfs have excessively large velocities, angular momenta and total energies to be long-lived, bound satellites of the MW. If that is the case, then the rotational support some MW satellites have is likely to be transient, increasing the likelihood of the arrangement being a chance occurrence. 

In this work, we start by first discussing the widely used `pole directions' approach, arguably introduced by \cite{kroupa2005}, expanded in \cite{pawlowski2013} to include kinematic considerations, and updated in \cite{fritz2018} and \cite{li2021} to include more position and velocity data with greater precision. The analysis utilises the directions of angular momenta from which a best-fitting plane is inferred. We refer to these analyses to make clearer the novel contributions of our own work. Our main objectives are (1) to introduce some novel tools for the analysis of plane structure in 3D space, key amongst which is `planarity', a new concept aimed at quantifying planar structure without recourse to the number or thickness of planes, and (2) to apply these tools to characterise MW satellite structure and its bearing on the accuracy of the $\rm{\Lambda CDM}$ paradigm. During the course of the latter, we show that the MW is supported by positional plane structure but that kinematic support cannot be convincingly demonstrated. Further, we apply these new tools to compare the expectations from the high-resolution \textsc{NewHorizon} cosmological hydrodynamical simulation \citep{2021dubois} with those calculated from MW observations, to show that positional planarity appears common in this high resolution simulation, in line with other recent work \cite[e.g.][]{2018welker,2020santos, 2021samuel}, and additionally that planarity in our mock galaxies is typically kinematically supported.

From here on the paper is structured as follows. In Sections \ref{prep} and \ref{stochastic} we describe the datasets used and explain how measurement uncertainty is quantified throughout the paper. In Section \ref{approach} we briefly describe the pole direction approach. We replicate some key results using our data, make note of some challenges and list additional desiderata for an alternative method. In Section \ref{alternative} we describe an alternative notion of planarity which focuses on quantifying to what extent satellite system positions and velocities can be explained by planes. In Sections \ref{application} and \ref{simulation} we apply our approach to the MW data and the \textsc{NewHorizon} simulation. We summarise our findings in Section \ref{summary}. 

\section{Datasets}
\label{prep}

Our analysis makes use of two datasets: (1) survey data from Gaia EDR3, and (2) \textsc{NewHorizon}, a high-resolution cosmological hydrodynamical simulation. 

\subsection{Gaia}

Gaia \citep{gaia2016} is a European space mission, designed to provide high quality astrometry, photometry, and spectroscopy of the stellar population of the MW environment, which includes the satellites of the MW. Here, we use the positional and velocity data from the EDR3 \citep{gaia2021} as prepared in \cite{li2021} and kindly provided by the authors. The data is a selection of known MW satellites from the literature \citep[e.g.][]{fritz2018,mcconn2020}, each with at least four spectroscopically measured stars in EDR3 from which galactocentric positions and velocity can be estimated. A total of 46 satellites are included in the analysis. \cite{li2021} take considerable care to account for both systematic and measurement uncertainties using a Bayesian approach. This may be viewed as an improvement over the data provided by \cite{mcconn2020}, which is similar but features smaller errors mainly due to the omission of systematic elements. \cite{li2021} provide  position and velocity vectors in Cartesian form, with the 16th, 50th and 84th percentile values provided for each component in each vector.

\subsection{\textsc{NewHorizon}}

\textsc{NewHorizon}\footnote{\href{http://new.horizon-simulation.org}
{http://new.horizon-simulation.org}} \citep{2021dubois} is a cosmological
hydrodynamical simulation which is a zoom-in of a region within the (142~Mpc)$^{3}$ volume Horizon-AGN
simulation \citep{dubois2014,2017kaviraj}. It utilises the \textsc{Ramses} adaptive mesh refinement code \citep{teyssier2002} with an effective grid resolution of $4096^3$. Initial conditions are generated according to a \textit{WMAP7} \citep{komatsu2011} cosmology
 ($\Omega_{\rm m}=0.272$, $\Omega_\Lambda=0.728$, $\sigma_8=0.81$, 
 $\Omega_{\rm b}=0.045$, $H_0=70.4 \ \rm km\,s^{-1}\, Mpc^{-1}$, and $n_s=0.967$). 
\textsc{NewHorizon} combines a contiguous 20~Mpc diameter spherical volume with extremely high stellar mass ($1.3\times10^{4}{\rm M_{\odot}}$) and spatial resolution ($\sim34$~pc), which allows us to resolve dwarf satellite galaxies with sizes comparable to the faintest known MW satellites. \textsc{NewHorizon} allows us to combine the high resolution of a zoom-in simulation with a contiguous region that is large enough to preserve large-scale cosmological structures at maximum resolution. This combination is essential in order to understand whether the planarity of the MW is reproducible in a $\rm{\Lambda CDM}$ Universe and -- if it is -- the potential formation mechanisms behind such planes.

AdaptaHOP \citep{2004aubert,Tweed2009} is applied to the \textsc{NewHorizon} stellar particle distribution to identify structures. A minimum structure size of 50 particles and a minimum density of 160 times the critical density is imposed. We measure the basic properties of each identified structure, namely: (1) $M_{\star}$ -- the total mass of star particles, (2) $X$ -- the centre of mass (barycentre) of star particles, and (3) $V$ -- the average velocity of star particles. Unless otherwise stated, we use the last time-step ($z\sim0.2$) in the simulation for comparisons. In this paper, we focus on galaxies with a mass similar to the MW ($M_{\star} > 10^{10}$ M$_{\odot}$). We require satellites to have $10^5$ M$_{\odot}$ < $M_{\star}$ < $10^{10}$ M$_{\odot}$ and to be within 3 to 45 effective radii of their host to keep the situation comparable to observations. For the MW this would include satellites within around 10 -- 160 kpc.

\section{Stochastic simulation methodology}
\label{stochastic}

The observational data describing the positions and velocities of MW satellites used throughout this work is subject to measurement errors which need to be taken into account in order to draw valid inferences. We make use of stochastic simulation as a means of accounting for the uncertainty in the positions and velocities of MW satellite galaxies in observational data. Due to the limits of our data, we treat errors in all vector components as independent, which could misrepresent error to some extent. We make use of the 16th, 50th and 84th percentiles provided in \cite{li2021} and compute an epsilon skew normal random variable \citep{mudholkar2000} for each vector component, which is a three parameter form of the normal distribution, allowing for it to be asymmetrically distributed around the mean. To produce a single replication of the dataset we independently sample all components from all vectors. In this analysis we primarily use replications to calculate the sample distribution for aggregate metrics such as Gini coefficient values. In our tests, using 1000 replications to produce statistics or histograms was adequate and using a larger number (for example 5000 replications) does not change the conclusions.

\section{The pole direction approach}
\label{approach}
In this section we will briefly review a widely used \citep{kroupa2005, pawlowski2013, fritz2018, li2021} approach in order to help clarify the contributions of our own tools in contrast. 

Galaxies on the same plane, orbiting in the same direction, will have the same direction of angular momenta (also known as `pole direction'). Pole direction is defined as the normalised cross product of the position and the velocity vector. We can calculate the normal vector for all satellites as a way of investigating whether there is a common plane which accommodates both their position and velocity vectors. Since planes (for each orbital direction) have unique normal vectors, we can proceed technically by looking for a vector which is within some acceptable tolerance of as many satellite pole directions as possible. The plane implied by the vector we find is the common, best-fitting plane.

Since pole direction vectors originate at zero, their direction can be described using 2 angles. This results in a simple 2D map which can be used to find clusters of pole directions. The common approach is to measure deviation in terms of cosine angle. \cite{pawlowski2013} use a tolerance of one `spherical standard distance' (around 29 degrees), while \cite{fritz2018} use 36.87 degrees. We use 36.9 degrees. Note, that a normal vector and its opposite (180 degree rotation) imply the same plane but the reverse orbital direction. We begin by reproducing the analysis with our data.

After projecting all galaxies to the angles indicated by their pole directions, we use a grid search with the objective of maximising the number of satellites within a tolerance of either the candidate vector or its flipped counterpart (i.e. the vector pointing in exactly the opposite direction) to find the common best-fitting plane. \cite{pawlowski2013} use the 8 most concentrated poles to find the common plane, \cite{fritz2018} use the plane location from \cite{pawlowski2013} and \cite{li2021} follow the methodology in \cite{fritz2018} closely.

\begin{figure}
  \centering
  \includegraphics[width=8.5cm]{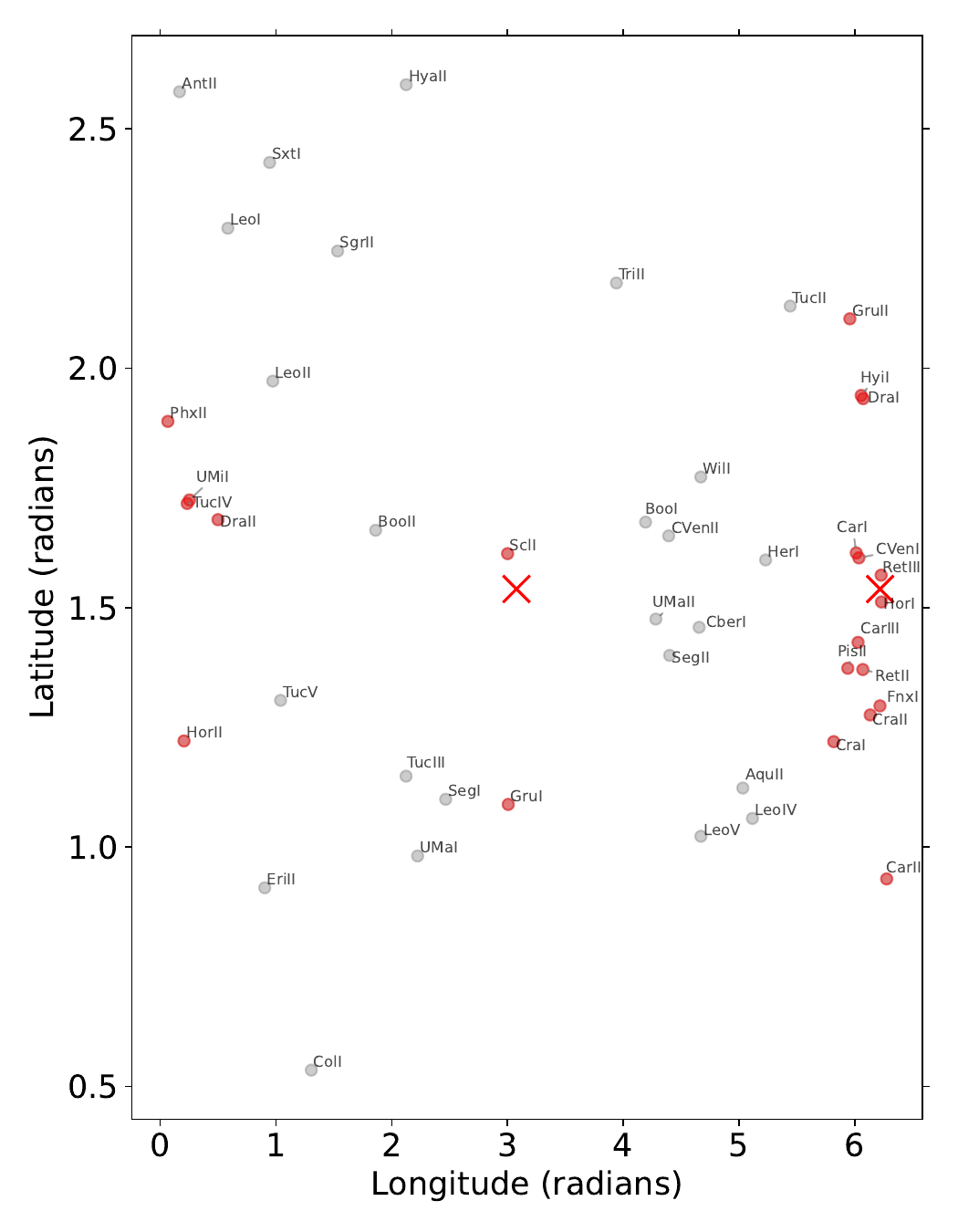}
  \caption {The pole direction of all 46 MW satellite galaxies, projected to spherical coordinates. The crosses show the two poles of the optimal normal vector corresponding to the plane which captures the most satellites. The satellites within 36.9 degrees of the plane are shown in red. There are 21 satellites within this tolerance, of which only 2 are counter-orbiting.}
  \label{poles}
\end{figure}

Figure \ref{poles} shows the pole direction of all 46 satellite galaxies projected to spherical coordinates. The crosses show the two poles of the optimal normal vector corresponding to the plane which captures the most satellites. The satellites within tolerance are shown in red. There are 21 satellites within 36.9 degrees of the plane, of which only 2 are counter-orbiting, which is very similar to the results published in \cite{li2021} using the same data but with a different method: they note that "20 [satellites] have median orbital poles that align to better than $\theta$ [36.9 degrees] with the VPOS
normal vector". The apparent single plane which accommodates 21 satellite galaxies lends credence to the existence of a major kinematically supported plane: the vast polar structure (VPOS) of satellite galaxies.\\

\subsection{Further requirements for our purposes}
\label{issues}

We intend to introduce a new abstract concept of `planarity' to quantify plane formation in satellite systems regardless of the number of planes, their thickness or location. The pole direction approach cannot be used for modelling this notion of planarity because it is primarily concerned with quantifying a single plane. We further think it would be useful to consider planes via positions and velocities as separate analyses, not least because errors associated with velocities are significantly greater, and therefore may obscure a picture otherwise clearer via positions only.

\begin{figure}
  \centering
  \includegraphics[width=8.5cm]{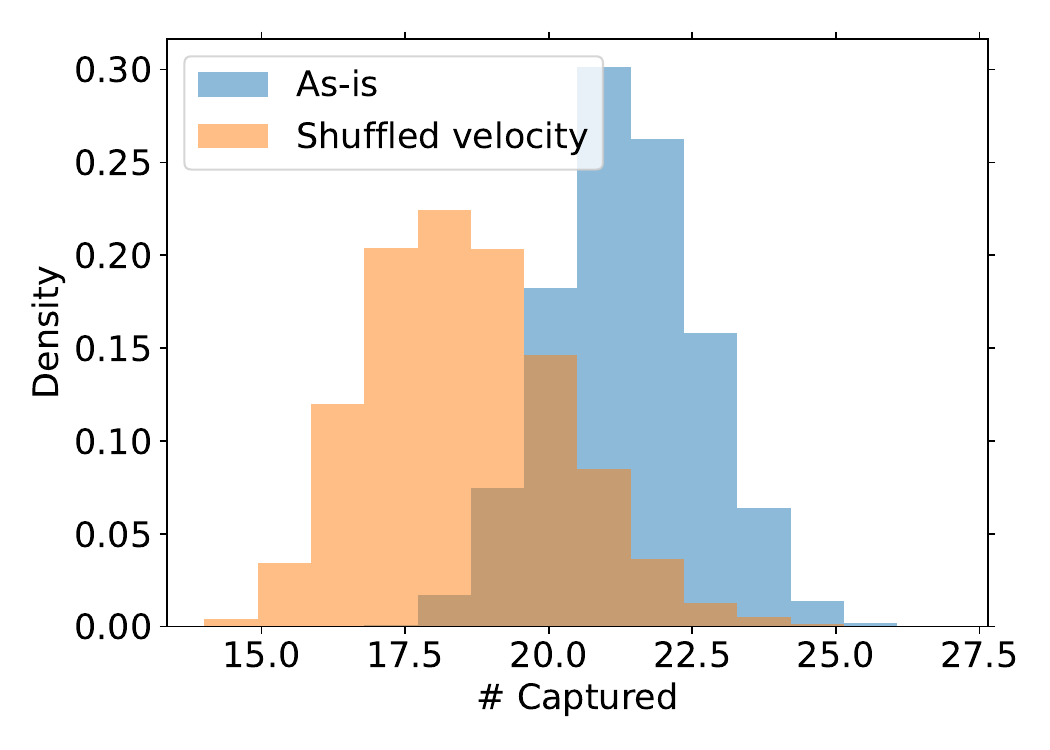}
  \caption {
    The maximal plane captured at a threshold of 36.9 degrees, produced by creating 1000 replications of positions and velocities for all satellites in two scenarios. In the `shuffled velocity' scenario, we shuffle the velocities between satellites and in the `as-is' scenario we make no changes. %The shuffled scenario averages about 18 captured satellites (the expectation would be $0.41 \times 46 \approx 19$) but extends to 22 within a 95 per cent confidence. Meanwhile, the as-is scenario averages about 22 captured satellites but extends to 24 within a 95 per cent confidence. 
    Around 35 per cent of the `as-is' distribution overlaps with the shuffled scenario. This implies a high chance of confusion with what is expected at random.
  }
  \label{shuffled}
\end{figure}

To illustrate the effect of the velocity errors, Figure \ref{shuffled} shows a histogram of the number of satellites on the biggest plane captured at a threshold of 36.9 degrees produced by creating 1000 replications of positions and velocities for all satellites in two scenarios. In the `shuffled velocity' scenario, we shuffle the velocities between satellites\footnote{Shuffling velocities breaks the connection between positions and velocities for any given satellite. The same result could be achieved by shuffling positions. The two procedures are equivalent.} and in the `as-is' scenario we make no changes. The shuffled scenario averages about 18 captured satellites but extends to 22 within a 95 per cent confidence. Meanwhile, the as-is scenario averages around 22 captured satellites but extends to 24 within a 95 per cent confidence. Around 35 per cent of the `As-is' distribution overlaps with the shuffled scenario implying a high chance of confusion with what may be expected at random. Had all satellites been on the same plane, shuffling would have no effect. We note, however, that just over half the satellites fall on a plane, therefore shuffling should have a pronounced effect. 

\section{The plane space approach}
\label{alternative}
We have stated in the previous section some additional desiderata. In this section we introduce some new tools which fulfil the desiderata and facilitate a more intuitive analysis for the study of planes in satellite systems. Our principal contribution is the \emph{plane space} which makes it easy to see every supported plane in a satellite system as an intuitive 2D image. This plane space can then be further summarised, making it possible to, for example, represent the \emph{planarity} of a satellite system -- the degree to which satellite positions and velocities are explained by planes -- in a single number, which is convenient for simulation and distribution building. 

We begin by describing how a plane space can be calculated and summarised in the case of satellite positions and then extend its application to velocities and angular momenta. We make the assumption that the planes in our analysis pass through the host. A 3D plane can be defined by three points, and therefore any two satellite position vectors combined with the origin point is enough information to uniquely define a 3D plane. When the plane passes through the origin, it can be defined with just two parameters: its angles in the spherical coordinate system. The angles defining the plane $\alpha,\beta$ (the inclination and azimuth respectively\footnote{Inclination is the angle made between the norm and the z-axis, whilst the azimuth is the angle made by the projection of the norm onto the xy-plane relative to the x-axis}) are given by: $\alpha = arccos(z/r), \beta = sign(y) * arccos(x/\sqrt{x^2+y^2})$ where $r=\sqrt{x^2 + y^2 + z^2}$ and $(x,y,z)$ is the cross product of the two position vectors. We proceed as follows:

\begin{enumerate}
    \item Using the equations above, for each unique pair of satellites in a satellite system around a host galaxy, we calculate their shared plane.
    \item We express the plane in spherical coordinates given by the 2 angles $\alpha$ and $\beta$. This represents one 2D point in the \emph{plane space}.
    \item We summarise the resulting 2D points from all pairs of satellites, into a $m \times m$ histogram, where $m$ represents the number of rows/columns (see below). This results in a binned count of planes implied by galaxy pairs.
\end{enumerate}

The histogram is binned such that a plane chosen at random from an isotropic distribution is equally likely to fall into any given cell. This can be achieved by dividing $\alpha$ linearly into $m$ equally sized bins, and by scaling $\beta$ by the cosine of $\alpha$ to account for the geometry of spherical space: $\bar{\beta} = \pi / (1 - cos(\alpha))$. This scheme results in a uniform distribution of counts in terms of either angle. Note that the plane space is built by considering every pair of satellites and fitting a plane to them. Therefore, if a plane captures at least two satellites, it will be counted in the histogram. The more common a plane is, the higher the count of satellite pairs which imply it, and the more concentrated the associated bin in the histogram. This gives the concentration of counts in the histogram a natural interpretation. Note also that since each plane has two possible orientations (identified by normal vectors pointing in opposing directions), each plane will fall into exactly two bins. This can be avoided by folding the space appropriately, but on balance we think that the simple treatment which admits the duplicates is more intuitive.

The crux of the plane space method is fitting a plane to three points in 3D space: one of them being the origin. The same method can be used verbatim with velocity vectors, even though they do not have the same origin. To see why, recall that velocity vectors indicate the rate of change in position across spatial dimensions, and imagine that they have the same origin (that is, imagine as if they were acting on the same object). It is then apparent that velocity vectors pointing along the same plane -- as would be revealed by their cross product -- must also be points falling on a common plane running through the origin -- as would be revealed above by fitting a plane to three points. Further it should be noted that if satellites are located on same plane that they are moving along, then their positional and velocity plane spaces will appear the same, enabling us to study kinematic support by comparing the plane spaces.

Another representation useful for the investigation of kinematic support is the plane space built using the cross product of satellite position and velocity vectors. If we again imagine these cross products to be rooted at the origin (as if acting on the same object), it is apparent that points falling onto a common plane running through the origin must also fall onto common position and velocity planes. The cross product of position and velocity vectors are angular momenta, so our method also extends to angular momentum vectors.

\begin{figure}
  \centering
  \includegraphics[width=4.2cm]{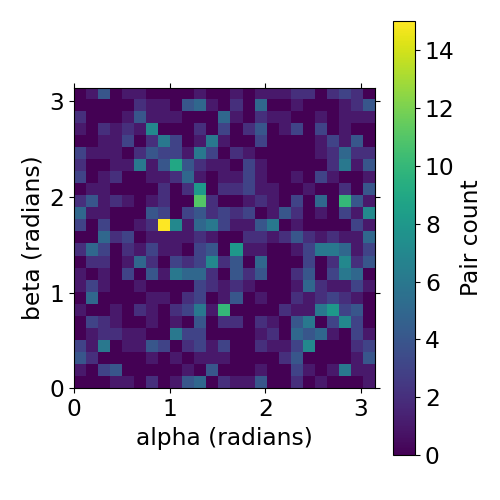}
  \includegraphics[width=4.2cm]{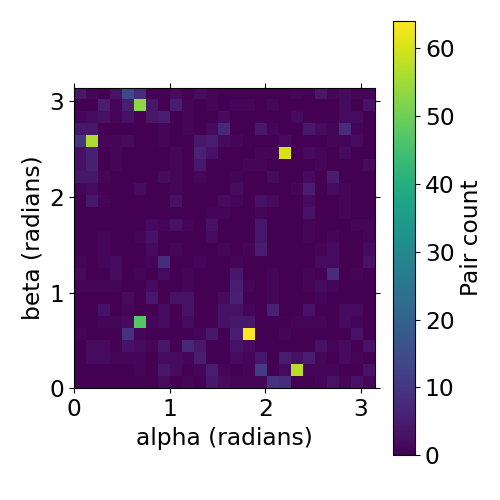}
  \includegraphics[width=4.2cm]{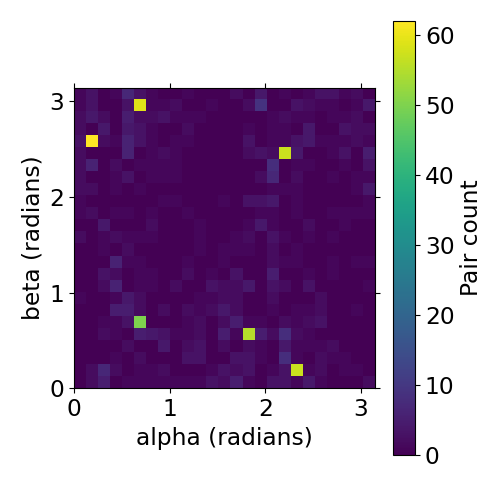}
  \includegraphics[width=4.2cm]{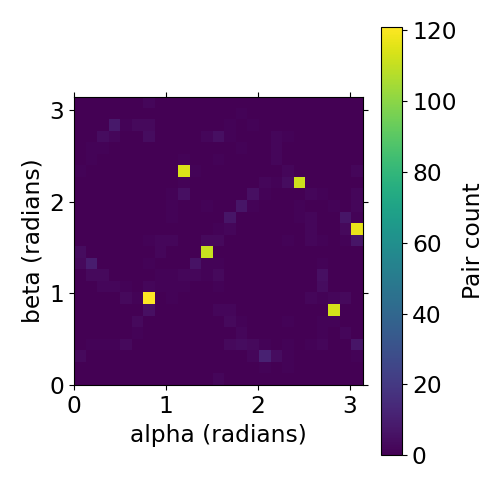}
  \caption {
    The top left panel shows the plane space for an isotropic distribution for position, velocity or angular momenta vectors. All three scenarios produce a uniform distribution across the histogram cells, but we present only one panel for brevity. The other panels illustrate what three discrete planes look like in plane space for positions (top right), velocities (bottom left) and angular momenta (bottom right) vectors respectively. The satellites in this example travel along their positional planes, so the position and velocity plane spaces look the same. Discrete planes result in concentration in the plane space.
  }
  \label{plane-space-vis}
\end{figure}

The top-left panel in Figure \ref{plane-space-vis} illustrates what the plane space looks like for an isotropic distribution for position, velocity or angular momenta vectors. That is, all three result in a uniform distribution across the histogram cells, with only incidental concentrations, but we present one panel for brevity. The other panels illustrate what three discrete planes look like in plane space for positions (top right), velocities (bottom left) and angular momenta (bottom right) vectors respectively. The satellites in this example travel along their positional planes so the position and velocity plane spaces look the same. There are six points (two per plane as explained above), all of which are highly concentrated, thus showing that discrete planes result in concentration in the plane space.

The plane space building procedure has one free parameter: $m$. The value of $m$ balances the resolution of the plane space with the sample size. The most important consideration is that `negative' scenarios (for example, randomly distributed satellites) are clearly separable from `positive' scenarios (for example, a high degree of planarity). We use $m=25$ -- a value calibrated by eye which adequately distinguishes between synthetic cases -- although in our experiments, the overall analysis is not sensitive to the precise value of $m$ (for example, $20<m<30$) in terms of its effect on summary measures used herein. We expect this is because small variations in bin width mostly, and at random, confound adjacent concentrations, which at fairly high resolutions ($m\ge 25$) have small individual contributions. That is, the 2D histogram is more likely to change gradually as resolution increases rather than change in big jumps between values of $m$.

\subsection{Using the plane space}

Producing intuitive 2D images is important because it facilitates informal hypothesis testing and exploratory analysis for the astronomer, but it leaves open questions like how a plane space should be interpreted for specific cases, what the thresholds are, and how plane spaces can be compared. In our analysis, we focus on `planarity', without attempting to determine the number of planes or plane specific properties. That is, `planarity' conceived as the amount of kinematic structure explained by planes. Figure \ref{plane-space-vis} shows that planarity is manifested as \textit{concentration} in the plane space, so we proceed by defining a single number summary statistic for describing this concentration. Such a metric would allow us to quantify how much planar structure exists in a satellite system without needing to concern ourselves with how many planes there are or the precise nature of the arrangement.

The Gini coefficient \citep{1936gini} is a metric well suited for summarising concentration. It has been extensively used in astronomy following its introduction in \cite{2004lotz} as a non-parametric approach to quantifying galaxy morphology. It ranges between 0 and 1, wherein higher values indicate higher levels of concentration. Our $m\times m$ histogram described earlier counts the occurrences of planes implied by pairs of galaxies along angular ranges in a spherical projection. We calculate the Gini coefficient as follows:
$$G = \frac{\sum_{i=1}^n \sum_{j=1}^n \mid x_i - x_j \mid}{2n^2\hat{x}}$$
where $x_\cdot$ are individual cell counts and $\hat{x}$ is the mean of the counts.

Since the number of cells in the histogram is fixed ($m^2$), whilst the number of planes represented may vary, the Gini coefficient will be sensitive to sample size. For example, if a host had two satellites, there would be just one possible plane, the whole plane space would be concentrated into two cells, and the Gini coefficient would be maximal, even if the satellite locations were selected at random. To remedy this issue, we will always use the Gini coefficient relative to a comparator distribution in the isotropic scenario with the same number of satellites as the host being analysed. This allows us to, for example, present the achieved Gini coefficient as a percentile of the isotropic distribution, or to concurrently present the isotropic baseline for ease of comparison, according to what is appropriate at the time.

For any given number of satellite galaxies, we establish the isotropic distribution by calculating the percentiles of the Gini index measured over 1000 random replications. The main desiderata is that the reference percentile of a Gini coefficient for an arbitrary plane space proportionally increases as the concentration of implied planes increases. That is, fewer planes must imply a higher percentile.

Note that the analysis herein is primarily concerned with quantifying planarity conceived as an aggregate structural measure as revealed by plane space concentrations through the Gini coefficient. To that end it is desirable that different spatial groupings in the plane space with the same Gini coefficients are considered as equivalent. However, this is not a universal prescription and other ways to summarise the plane space may be appropriate for other kinds of analyses. For example, using the histogram representation directly is well suited to machine learning tasks such as the clustering or classification of different kinds of planar arrangements.

Note also that our analysis assumes that planes pass through the origin (host). We think this is a sensible assumption in most cases because the dark matter halo of the host galaxy encompasses both the host and its satellites, and its centre is usually assumed to be at or near the host galaxy, although some simulation analyses \citep[e.g.][]{2023Santos} suggest that the distance from the centre to the best fitting plane(s) may vary over cosmic time. In the latter case, our method will still capture structure, but a large offset from the centre may lead to a lower Gini coefficient being attributed to the satellite system. Since we see large Gini coefficients in simulation, our assumption seems reasonable, but we nonetheless present it here as a possible concern.

We complete this section by briefly reminding the reader how our plane space and Gini coefficient approach compares to previous measures in the literature: pole direction \citep{kroupa2005, pawlowski2013, fritz2018, li2021}, root-mean-square height \citep{samuel2021, pawlowski2013}, minor-to-major axis ratio \citep{samuel2021}. As detailed in Section \ref{approach}, pole directions can be used to project galaxies into a 2D angular space. Spatial clustering techniques can then utilised to hunt for planes. A challenge is that using pole directions can be sensitive to measurement errors resulting in spurious coherence (see Figure \ref{shuffled}). In our method, positions and velocities can be analysed separately.

Root-mean-square (RMS) height (perpendicular distance from a satellite to a plane) is typically used to measure plane thickness. It can be used to qualify planes as sufficiently `thin' whilst some other method is used to generate planes to be tested \citep{samuel2021}. The closest analogue to `thickness' in our method is the parameter $m$ which serves to constrain the enclosing angle of any given plane, however it is not a direct replacement for it.

The minor-to-major axis ratio decomposes the inertial tensors of each satellite into eigenvalues and the takes the ratio of the square roots of the smallest and biggest value. A ratio close to 1 indicates an isotropic distribution whilst values closer to zero represent a flat spatial distribution. Unlike our approach, this measure is limited to single plane systems.

Finally, we note again that generally other treatments of planes in the literature are typically focused on quantifying a single plane as opposed to `planarity' as described herein, so to some extent these comparisons are not like-for-like.

\section{Application to MW satellites}\label{application}

We now apply the tools described in the previous sections to the MW. We begin by building up a picture of the MW plane structure via position and velocity vectors. 

\begin{figure}
  \centering
  \includegraphics[width=6.5cm]{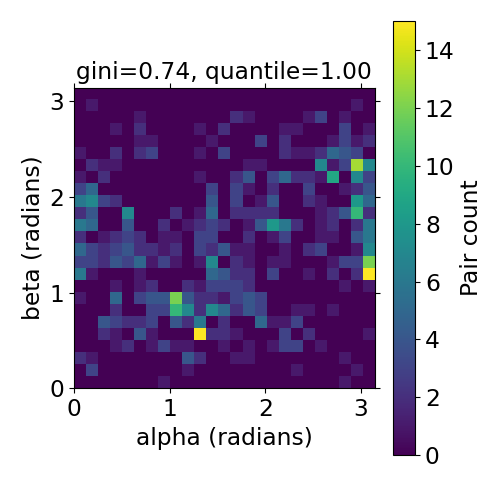}
  \includegraphics[width=6.5cm]{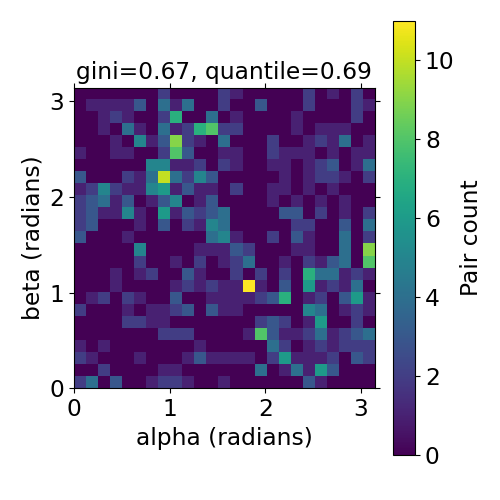}
  \caption {
    The MW plane space constructed using position (top) and velocity (bottom) vectors respectively. The colours highlight the number of galaxy pairs falling into any given bin. The title indicates the Gini coefficient and the quantile of the isotropic distribution that the value falls in. Compared to the isotropic satellite distribution, only the position vectors provide a compelling case, whilst the velocity vectors have a statistically insignificant Gini coefficient at a 95 per cent confidence.
  }
  \label{mw}
\end{figure}

From top to bottom, the panels in Figure \ref{mw} show the MW plane space constructed using median position and velocity vectors respectively. The colours highlight the number of galaxy pairs falling into any given bin. The titles indicate the Gini coefficient and the quantile of the isotropic distribution that the Gini value falls in. If prominent planes exist in the MW, we would expect to see concentration in the plane spaces. Since the positions have the lowest measurement errors, we would expect planes to be most clearly visible in the position plane space. We note that these images do not take into account measurement variance and a coherent looking median image by itself is not enough to establish planarity.

In position plane space, we can see concentrations, and it is confirmed by a Gini value in the 100th percentile. We note that the median velocity plane space does not visually correspond to the position plane space and that the concentrations exhibited are likely random according to the Gini coefficient. This non-correspondence of the median images is already enough to raise intuitive suspicion. If planes are kinematically supported (that is, if satellite velocities are coherent with satellite positions), we would expect the position and velocity spaces to overlap visually -- at least in part, and we would expect the median images in both cases to be concentrated to a statistically significant degree.

\begin{figure}
  \centering
  \includegraphics[width=8.5cm]{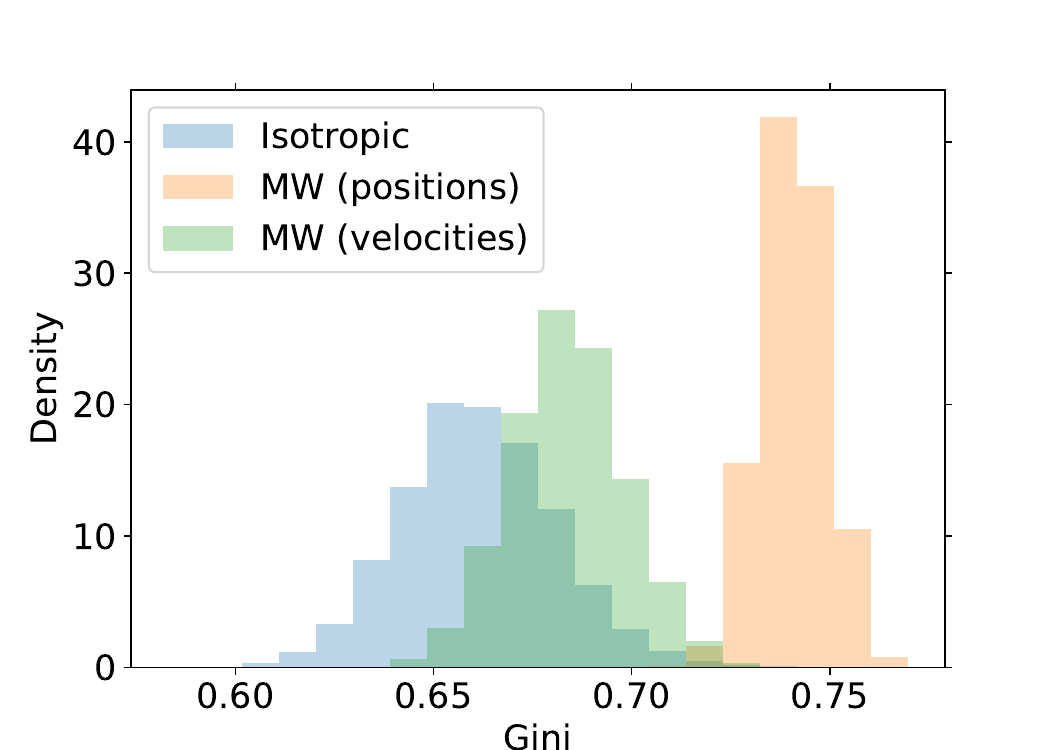}
  \caption {
    Histograms created from 1000 replications, drawn from the isotropic, positional and velocity vector distributions of the MW are shown over-plotted. Given the measurement errors, only the positional plane space is significantly differentiated from the isotropic distribution.
  }
  \label{dist-comparison}
\end{figure}

To better quantify the planarity and take into account measurement error, we create 1000 replications (see Section \ref{stochastic}) from position/velocity measurements. Figure \ref{dist-comparison} over-plots histograms created from 1000 replications, drawn from the isotropic, positional, velocity and angular momenta vector distributions respectively. The diagram makes clearer both the relative offsets of the distributions from the isotropic case, and also the relative variance in each case. Only the positional plane space is significantly differentiated from the isotropic distribution: more than 99 per cent of the position vector replications and only 20 per cent of the velocity vector replications have a Gini coefficient greater than the 95th percentile of the isotropic distribution. Note also the large relative variance of the velocity distribution in comparison to the positional distribution, which may explain the lack of correspondence, even if it did physically exist.

To determine kinematic coherence, we expect to find support for it in both the position and velocity plane space independently. However, we conclude that significant planarity is supported by the position plane space but not by the velocity plane space. Comparison of the median images do not suggest the same pattern of concentration but it is possible that the relatively high measurement error in the velocity vectors is obscuring a correspondence.

We have mentioned in Section \ref{approach} that it is a common conclusion \citep[e.g.][]{kroupa2005, pawlowski2013, fritz2018, li2021} that the MW has at least one prominent and kinematically supported plane (VPOS). We have also described in Section \ref{approach} why the conclusion may be confounded. We further note that there are others who have not been able to replicate related results. For example, \cite{2023Till} use Gaia data and dynamically model the MW arguing that it is unlikely that the MW satellite plane is rotationally supported and that alignments are more likely to be transient. Further, although \cite{2012Pawlowski2} claimed that the positions of young globular clusters and stellar/gaseous streams aligned with the VPOS plane, \cite{2020Riley} -- using more accurate data from Gaia -- were not able to replicate this, asserting that they do not align. Our conclusions using the plane space method are therefore in agreement with similar conclusions arrived at in other ways.

While, for the purposes of the work herein, establishing a planarity baseline in the MW is enough, interested readers could take this analysis further. For example, the plane space can also be used to determine which galaxies least contribute to space concentration by leaving each galaxy out in turn, recalculating the plane space and its Gini coefficient, and then comparing how much the Gini coefficient has changed. This ranking can, in turn, be used with satellite galaxy properties, such as the distance from the host, mass and other measured quantities, to better understand the covariates of planarity.

\section{Application to the \textsc{NewHorizon} simulation}\label{simulation}

In this section, we compare the key results from the MW plane spaces, with realisations from the NewHorizon cosmological hydrodynamical simulation, in order to investigate whether the planarity observed in the MW is unusual in the context of the $\rm{\Lambda CDM}$ model, even if kinematic support is assumed.

\subsection{Comparing the MW to NewHorizon}

In order to keep the simulated host galaxies comparable to the MW, we consider NewHorizon hosts with stellar masses greater than 10$^{10}$ M$_{\odot}$, which have more than 30 satellite galaxies at the final time step and where each satellite galaxy has a stellar mass greater than 10$^{5}$ M$_{\odot}$. This results in around 20 eligible host galaxies at the final time step of the simulation ($z\sim0.2$) and a varying number of eligible hosts at other time steps.

\begin{figure}
  \centering
  \includegraphics[width=8.5cm]{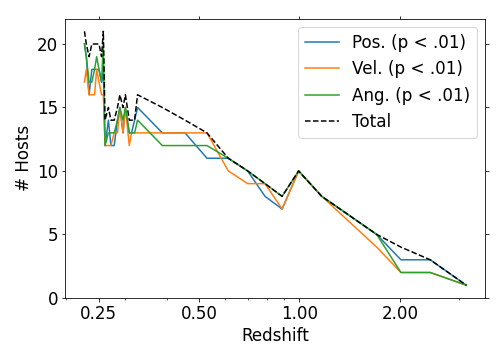}
  \caption {
  The total number of host galaxies at each time step (black), and the number of galaxies with a Gini coefficient in the 100th percentile of the isotropic distribution (equivalent to the MW) for positions (blue), velocities (orange) and angular momenta (green). At least 90 per cent of host galaxies exhibit positional planarity at a level comparable to or greater than the MW. Furthermore, positions, velocities and angular momenta are highly correlated, demonstrating kinematic coherence.
  }
  \label{long-all}
\end{figure}

We begin by considering planarity, as proxied by the Gini reference percentiles, for all eligible galaxies at every time step in the redshift range $0.2<z<2.5$. In the analysis to follow, we make use of planarity calculated from position, velocity and angular momenta spaces separately. Note that our simulations do not have measurement errors, so angular momenta have been included to cross-validate that effects measured in position and velocity plane spaces have coherent implications for angular momenta.

Figure \ref{long-all} shows the total number of host galaxies at each time step (black), and the number of galaxies with a Gini coefficient in the 100th percentile (equivalent to the MW, see Section \ref{application}) of the isotropic distribution for positions (blue) and velocities (orange). At least 90 per cent of MW-like host galaxies exhibit positional planarity at a level comparable to or greater than the MW. It is also notable that positions, velocities and angular momenta are highly correlated, demonstrating kinematic coherence. This strongly suggests that significant planarity amongst MW-like massive hosts is common in the simulation, and that the planes tend to be kinematically supported. It also suggests that this phenomenon exists largely independent of cosmic time.

\begin{figure}
  \centering
  \includegraphics[width=8.5cm]{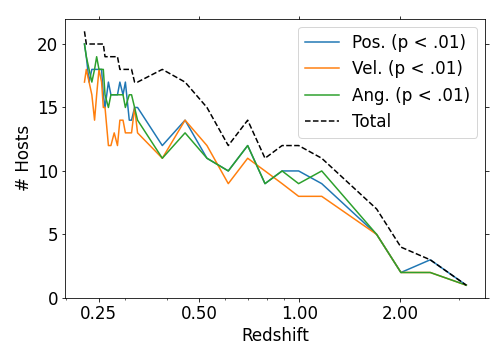}
  \caption {
  The total number of host galaxies at each time step (black), and the number of galaxies with a Gini coefficient in the 100th percentile of the isotropic distribution for positions (blue), velocities (orange) and angular momenta (green). The plot is filtered to include just those galaxies eligible at the final time step. At least 80 per cent of host galaxies exhibit positional planarity, at a level comparable to or greater than the MW, throughout their lifetime.
  }
  \label{949-long-summary}  
\end{figure}

In Figure \ref{949-long-summary}, we further consider whether the same host galaxies tend to remain highly planar over time or whether it is a transient phenomenon. Similarly to Figure \ref{long-all}, we show the total number of host galaxies at each time step (black), and the number of galaxies with a Gini coefficient in the 100th percentile of the isotropic distribution for positions (blue) and velocities (orange). However, this plot shows just those galaxies that are eligible hosts at the final time step, therefore providing a view of their planar structures over cosmic time. At least 80 per cent of these host galaxies exhibit positional planarity, at a level comparable to or greater than the MW, throughout their lifetime. This suggests that planarity in massive hosts is a lifelong feature in the simulation. It also further confirms that positions, velocities and angular momenta are highly correlated at all times, demonstrating kinematic coherence.

\begin{figure}
  \centering
  \includegraphics[width=8.5cm]{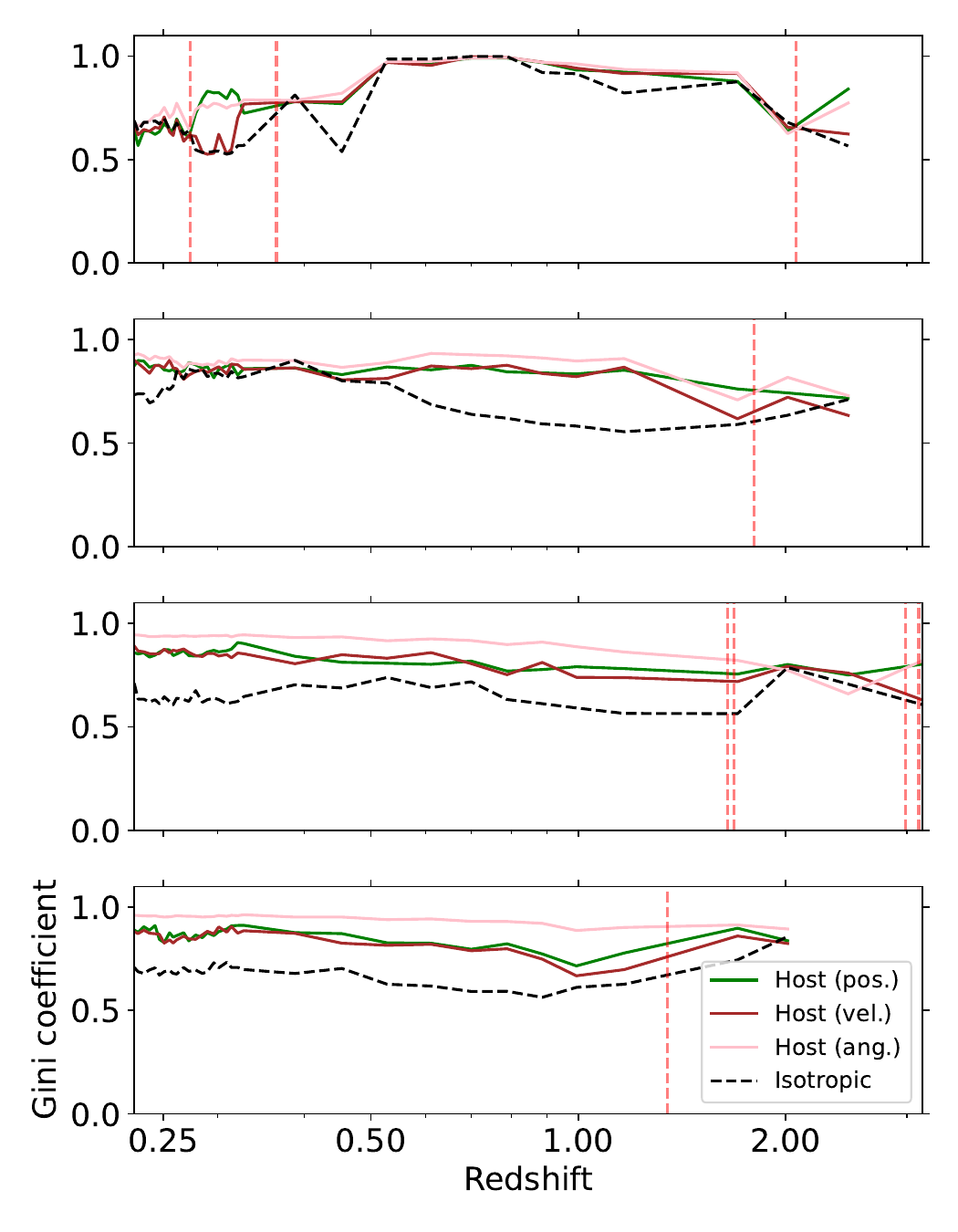}
  \caption {
  Evolution of the Gini coefficient for four example host galaxies for positions (green) and velocities (brown). The dashed black line shows the 100th percentile threshold which represents the MW equivalent position Gini coefficient. The vertical lines show minor/major mergers. The black line changes because the number of satellites in the sample changes over time. Hosts tend to start and remain above the threshold, with positions, velocities and angular momenta tracking each other and exhibiting similar levels of planarity. Minor/major mergers are uncorrelated with planarity.}
  \label{949-long}
\end{figure}
 
We confirm this further by manually checking the evolution of the Gini coefficient for all eligible hosts at the last time step. In Figure \ref{949-long} we show the evolution of the Gini coefficient for four example host galaxies (the trends are similar in most other hosts, which are not shown for brevity). The dashed red line shows the 100th percentile threshold, which represents MW equivalent planarity. The dashed black line shows the 100th percentile threshold which represents the MW equivalent position Gini coefficient. The vertical lines show minor and major mergers during the period. Here minor mergers are defined as those with mass ratios between 1:10 and 1:4, while major mergers have mass ratios greater than 1:4. The black line changes because the number of satellites in the sample changes over time. In these examples, and generally, eligible hosts tend to start and remain above the threshold. Positions, velocities and angular momenta tend to track each other and therefore suggest kinematic coherence. We observe that planarity varies over time -- this is mostly because satellite membership varies over time causing planar arrangements to change -- but it tends to start and remain above the threshold, indicating that, although individual planes could be short-lived, kinematically supported planarity remains constantly present. We also observe that minor/major mergers appear uncorrelated with planarity, although it is worth noting that around 90 per cent of such mergers for most eligible galaxies occur at redshifts greater than 3.

If these observations were characteristic of MW sized galaxies, we would hypothesise that kinematically coherent planes must be a feature of structure formation, in favour of which there is some observational precedence \citep{2008paz,2015tempel} as well as substantial argument from first principles. $\rm{\Lambda CDM}$ implies hierarchical structure formation wherein smaller structures form first and later merge to form larger ones. This process is directional due to the cosmic web structure of the universe: the large-scale structures including galaxy clusters and filaments created by the gravitational attraction of dark matter and gas. The filamentary structure of the cosmic web may imply that dwarf galaxies or proto-galactic fragments tend to fall into larger galaxies along specific directions; not randomly \citep[e.g.][]{2004aubert,2008paz,2011pichon, 2012codis, 2015libeskind,2015buck,2015tempel}. This could lead to the initial planar distribution of satellite galaxies around hosts. Further, once a planar structure has formed, dynamical friction, and the conservation of angular momentum in the host's dark matter halo may work to maintain this structure and create velocity coherence among the satellites by drawing satellites closer to the host (thus reducing the potential number of trajectories), and by forcing satellite orbits to synchronise with the rest of the matter in the halo \citep{2018welker}.

As discussed in Section \ref{application}, positional planarity is significantly non-random in the MW but velocity related measurement errors are too high to confirm kinematic support. However, if the simulation results are indicative of MW-like galaxies, we can expect that as MW satellite galaxy velocity measurements improve, kinematic support may be confirmed.

In the simulation, we also observe a significant weakening of planarity (1) amongst hosts at sub MW stellar mass, and (2) when the minimum stellar mass of satellites is increased ($M_{\star} > 10^{7}$ M$_{\odot}$). This could have many possible explanations but it is consistent with the hypothesis above, since in both cases, the stellar mass difference between the host and the satellite is reduced, and relatively less massive hosts are more significantly affected by their environment. That is, even if satellites were initially torqued into planar orbits, their own weaker potential may imply that the plane is more likely to be disrupted by external perturbations (such as the overall tidal field of the galaxy group, close interactions, and so on), which could wash out the initial configuration.

\section{Summary}\label{summary}

This paper had the following objectives: (1) to introduce novel tools which enable a more descriptive analysis of planarity and (2) to apply these tools to characterise the MW satellite structure and to explore whether the observed planar structures are in tension with simulated galaxies in the $\rm{\Lambda CDM}$ paradigm. We motivated the new tools on the basis that they are more geared towards representing an abstract notion of planarity, which does not require commitment to the number of planes or their thickness, and factoring spaces (positions and velocities) that may have been considered together in past analyses. We have used these tools to produce an analysis of MW planarity and to compare the results with NewHorizon -- a $\rm{\Lambda CDM}$-based high-resolution cosmological hydrodynamical simulation -- to assess whether observations are generally in agreement with currently accepted theory. 

Our tools use a simple procedure to derive a \emph{plane space} which describes all supported planes of a satellite system. The same method is applicable to analysing positions, velocities and angular momenta. We have shown that this plane space may be summarised in various ways and that it can be used in stochastic simulation. We have applied these tools to the study of MW planarity and then compared the results to a similar analysis derived from MW type galaxies extracted from NewHorizon. Our main conclusions are as follows: 

\begin{itemize}

\item \emph{The positional distribution of MW satellites is significantly planar compared to an isotropic satellite distribution}. Clear concentrations are visible in its plane space image and the corresponding Gini coefficient is well separated from an isotropic comparator to a statistically significant level. However, the velocity plane space is not planar to a statistically significant level -- possibly due to high measurement errors -- and therefore it cannot be concluded, from the data available, that MW planarity is kinematically supported.

\item \emph{The NewHorizon simulation suggests that a MW level of planarity is common in MW-like simulated galaxies, and that planes are kinematically supported}. Considered both at individual time steps and across cosmic time, MW-like simulated host galaxies tend to be similarly planar, and they tend to maintain their planarity over cosmic time. The merger history i.e. the incidence of minor/major mergers do not appear correlated with planarity. 
\end{itemize}

We therefore conclude that the MW has significant positional plane structure, and that it is coherent with the $\rm{\Lambda CDM}$ paradigm. Finally, we hypothesised that the ubiquity of planarity in our sample suggests that it is made likely by the nature of hierarchical structure formation; possibly because the filamentary structure of the cosmic web implies that dwarf galaxies or proto-galactic fragments tend to fall into larger galaxies along specific directions, after which dynamical friction and angular momentum conservation contribute to the maintenance of kinematically-supported planes. Given this, we expect that higher precision measurements of MW satellite velocities may reveal kinematic support of its positional planes.

\section*{Acknowledgements}
We kindly thank the referee for many constructive comments that improved the quality of the original manuscript. We thank the authors of \cite{li2021} who provided pre-processed data from EDR3 from our analysis. EU acknowledges a PhD studentship from the Centre for Astrophysics Research at the University of Hertfordshire. SK acknowledges support from the STFC [grant number ST/X001318/1] and a Senior Research Fellowship from Worcester College Oxford.

This work has made use of data from the European Space Agency (ESA) mission Gaia (https://www.cosmos.esa.int/gaia), processed by he Gaia Data Processing and Analysis Consortium  DPAC, https://www.cosmos.esa.int/web/gaia/dpac/consortium). It has also made use of the Horizon cluster on which the simulation was post-processed, hosted by the Institut d'Astrophysique de Paris. We warmly thank S.~Rouberol for running it smoothly. This work is partially supported by the grant
Segal ANR-19-CE31-0017 of the French Agence Nationale de la Recherche and by the National Science Foundation under Grant No. NSF PHY-1748958.
This work was granted access to the HPC resources of CINES under the allocations  c2016047637, A0020407637 and A0070402192 by Genci, KSC-2017-G2-0003, KSC-2020-CRE-0055 and KSC-2020-CRE-0280 by KISTI, and as a “Grand Challenge” project granted by GENCI on the AMD Rome extension of the Joliot Curie supercomputer at TGCC. The large data transfer was supported by KREONET which is managed and operated by KISTI.  S.K.Y. acknowledges support from the Korean National Research Foundation (NRF-2020R1A2C3003769).  This study was funded in part by the NRF-2022R1A6A1A03053472 grant.

For the purpose of open access, the authors have applied a Creative Commons Attribution (CC BY) licence to any Author Accepted Manuscript version arising from this submission.

\section*{Data availability}\label{avail}

\noindent Code for all results and figures can be found at
\url{https://emiruz.com/vpos}.

\bibliographystyle{mnras}
\bibliography{paper}

\end{document}